\documentclass{jpp}
\usepackage{graphicx}
\usepackage[siunitx]{circuitikz}
\usepackage[utf8]{inputenc}
\usepackage[T1]{fontenc}
\usepackage{amsmath}

\shorttitle{Slow positron production and storage for the ASACUSA-Cusp experiment}
\shortauthor{D.~J.~Murtagh on behalf of the ASACUSA collaboration}

\title{Slow positron production and storage for the ASACUSA-Cusp experiment}

\author{
D.~J.~Murtagh \aff{1} \corresp{\email{daniel.murtagh@oeaw.ac.at}},
C.~Amsler \aff{1},
H.~Breuker \aff{2},  
M.~Bumbar \aff{1},
S. Chesnevskaya \aff{1},
G. Costantini \aff{3},
R. Ferragut \aff{4}, 
M. Giammarchi \aff{5},
A. Gligorova \aff{1}, 
G. Gosta \aff{3}, 
H.~Higaki \aff{6}, 
E.~D.~Hunter \aff{1},
C.~Killian \aff{1}, 
V.~Kraxberger \aff{1},
N.~Kuroda\aff{8}, 
A.~Lanz \aff{1},
M.~Leali \aff{3},
G.~Maero \aff{9},
C.~Mal\-bru\-not\aff{10}\footnote{present address: TRIUMF, Vancouver, Canada}, 
V.~Mascagna \aff{3},
Y.~Matsuda\aff{8},  
V.~M{\"a}ckel$^{a}$\footnote{present address: INFICON GmbH, K\"oln},
S.~Migliorati \aff{3}, 
A.~Nanda \aff{1},  
L.~Nowak\aff{1,10},
F.~Parnefjord~Gustafsson\aff{1,10},
S.~Rheinfrank \aff{1},
M.~Romé \aff{9},
M.~C.~Simon \aff{1},
M.~Tajima\aff{7},  
V. Toso \aff{4}, 
S.~Ulmer \aff{2},  
L.~Venturelli \aff{3},
A.~Weiser \aff{1},
E.~Widmann\aff{1},
T.~Wolz\aff{10},
Y.~Yamazaki \aff{2}, 
\and 
J.~Zmeskal\aff{1}.
}

\affiliation{
\aff{1}Stefan Meyer Institute for Subatomic Physics, OEAW, %%Vienna,
\aff{2}Ulmer Fundamental Symmetries Laboratory, RIKEN,
\aff{3}Dipartimento di Ingegneria dell'In\-formazione, Universit\`a degli Studi di Brescia, %%Brescia, Italy
and INFN Pavia, %Istituto Nazionale di Fisica Nucleare (INFN), %%Italy
\aff{4}Politechnico di Milano,
\aff{5}INFN Milano,
\aff{6}Graduate School of Advanced Science and Engineering, Hiroshima University,
\aff{7}Nishina Center for Accelerator-Based Science, RIKEN,
\aff{8}Institute of Physics, the University of Tokyo,
\aff{9}Dipartimento di Fisica, Università degli Studi di Milano and INFN Milano,
\aff{10}Experimental Physics Department, CERN.
}   

\begin{document}

\maketitle
\begin{abstract}
The ASACUSA Cusp experiment requires the production of dense positron plasmas with a high repetition rate to produce a beam of antihydrogen. In this work, details of the positron production apparatus used for the first observation of the antihydrogen beam, and subsequent measurements are described in detail. This apparatus replaced the previous compact trap design resulting in an improvement in positron accumulation by a factor of ($52\pm3)$.  
\end{abstract}

\section{Introduction} 
The ASACUSA Cusp experiment aims to perform spectroscopy of the hyperfine structure (HFS) of ground state antihydrogen in a magnetic field free region using a Rabi spectroscopy method (\cite{widmann_measurement_2004}). This experiment requires a spin polarised beam of ground state antihydrogen which is produced by mixing positrons ($\textrm{e}^+$) and antiprotons ($\bar{\textrm{p}}$) within the so-called Cusp trap (\cite{mohri_possible_2003}). Ground state antihydrogen ($\bar{\textrm{H}}$) produced in the Cusp trap with low enough velocity will be spin polarised and focused on axis, thus producing the beam required for spectroscopy (\cite{nagata_development_2015}). Antihydrogen has already been observed 2.7~m downstream of the production region (\cite{kuroda_source_2014}) and the initial population of the antihydrogen levels measured (\cite{kolbinger_measurement_2021}). For efficient production of antihydrogen, positrons need to be readily available for mixing with antiprotons.\par
The positron source and trap used for the first observation of antihydrogen by the ASACUSA collaboration (\cite{enomoto_synthesis_2010}) was documented in \cite{imao_positron_2010}, this arrangement used a slow beam generated from a tungsten foil moderator, and a 40cm long gas cell to trap positrons. This setup was replaced with a beam derived from a neon moderator and a new trap design in 2011 which has been described briefly elsewhere (e.g. \cite{kuroda_source_2014}). The production and storage of positrons are the subject of this work, for the first time a complete description of the system will be given and discussed.\par
\section{Slow beam production} 
Positrons are produced by a $^{22}\textrm{Na}$ source supplied from iThemba labs with the activity of 1.87~GBq in 2013. These fast positrons are slowed down using a neon moderator (\cite{mills_solid_1986}) grown onto a cone structure located directly in front of the source. The moderator arrangement was previously used at RIKEN (e.g. \cite{oshima_new_2000}), however the source was replaced upon its arrival at CERN. 

\begin{figure}
  \centering
  \includegraphics[scale = 0.3]{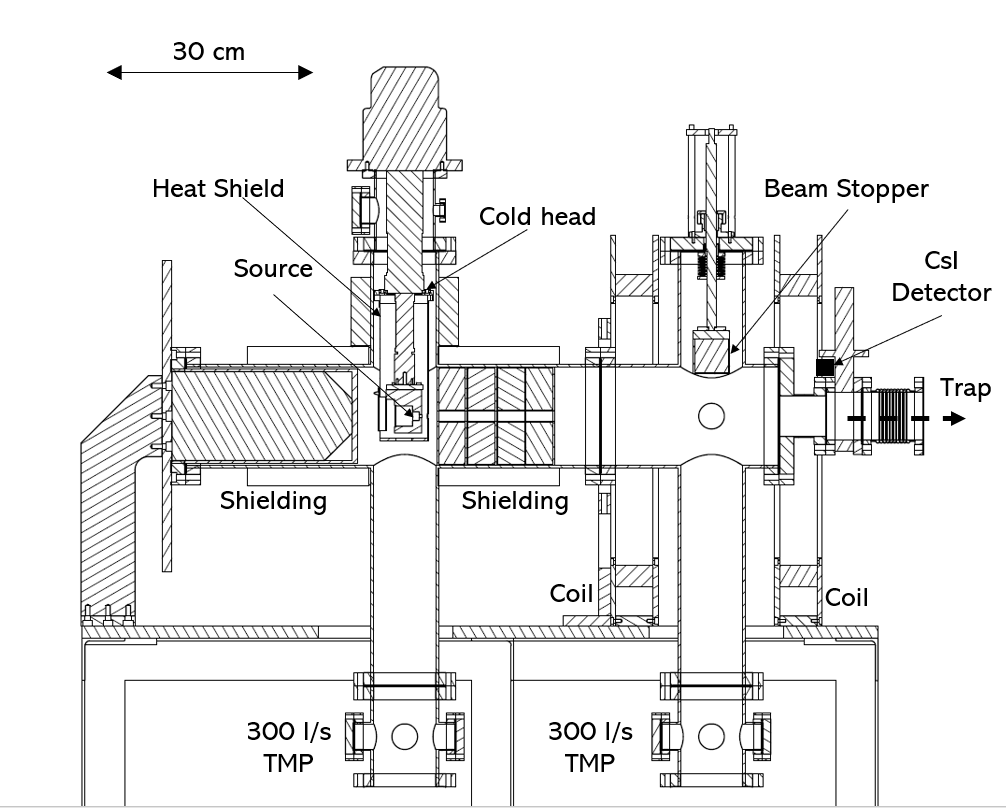}
  \caption{Scale drawing of the apparatus used for slow beam production.}
\label{fig:RGM}
\end{figure}
 
Figure \ref{fig:RGM} shows a scale diagram of a slice through the slow beam production apparatus. Inside the vacuum system, biological shielding close to the source is provided by lead sealed inside aluminium pots. The narrow aperture (15~mm diameter and 120~mm long) also serves to provide a pumping restriction between the moderator chamber and the downstream vacuum. There is a retractable beam blocker mounted on a linear manipulator which is operated by an external pneumatic actuator to stop the beam. This removes any direct line of sight to the source during interventions. A large external shielding made of 316 stainless steel is mounted inside a ``top hat'' style chamber\footnote{i.e. a chamber with a large internal cavity where the flange would be the brim of the so called ``top hat''.} outside of the vacuum which is also visible in figure \ref{fig:RGM} to the rear of the source. The vacuum chamber is surrounded by 6 water-cooled magnetic field coils (not shown in figure \ref{fig:RGM}) which can produce fields of up to 500~gauss (at 60~A) around the source. Finally the system is surrounded by lead bricks to bring the surface dose rate to 1-3 $ \mu \textrm{Svh}^{-1}$ at full source activity (1.87~GBq).\par
The moderation efficiency ($\epsilon_m$) of the neon ice is: 
\begin{equation}
    \epsilon_m = \frac{N^+}{A}
\end{equation}
where $N^+$ is the number of positrons in the slow beam and $A$ is the source activity in Bq. With a fresh uncontaminated moderator and a 1.87~GBq source the slow beam intensity is approximately 5 million per second. This is measured using a small CsI detector from which the count rate is corrected for attenuation and solid angle (position indicated in figure \ref{fig:RGM}). We have measured a moderation efficiency of $\epsilon_m = (0.25 \pm 0.1)$~\%. A new moderator is grown once per day due to a rapid decay rate measured to be ($16.6 \pm 3.4$)\%h$^{-1}$, this rate is due to contamination from buffer gases used in the trap discussed below.  

\section{Positron trapping} 
\begin{figure}
  \centering
  \includegraphics[scale = 0.3]{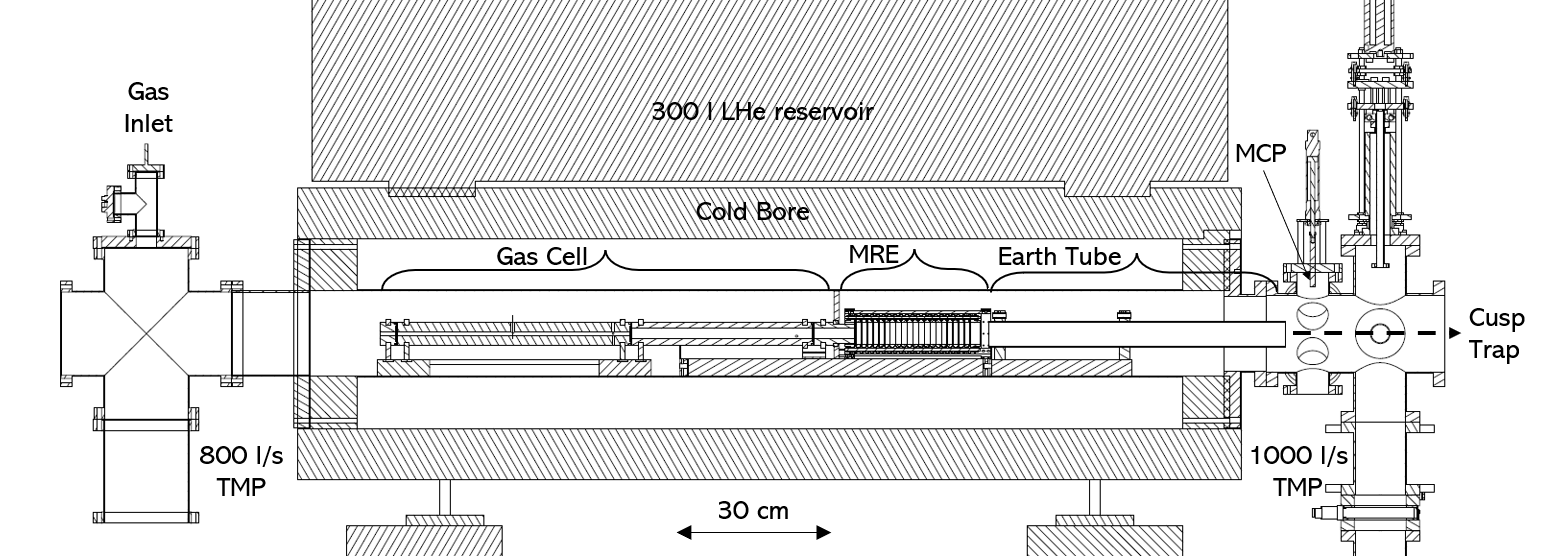}
  \caption{Scale drawing of the positron trap system.}
\label{fig:trap}
\end{figure}

The slow beam produced by the rare gas moderator is magnetically guided from the source into the trap magnet (centre of figure \ref{fig:trap}). The trap is housed within the cold bore of a superconducting magnet which is energised to 6000~gauss. The liquid helium reservoir (LHe) and the cold bore are in close contact, hence, it is not possible to run with a room temperature bore. As the bore will heat the reservoir this leads to an extremely high LHe consumption rate ($\sim$~300~l per 24~h at room temperature). The temperature is lowered to around 100~K using LHe cooling loops at either end of the vacuum tube. The trap is in contact with the cold bore, although no special attention was given during design to ensure good thermal conduction, nor is the temperature monitored.\par 
\begin{figure}
  \centering
  \includegraphics[scale = 0.3]{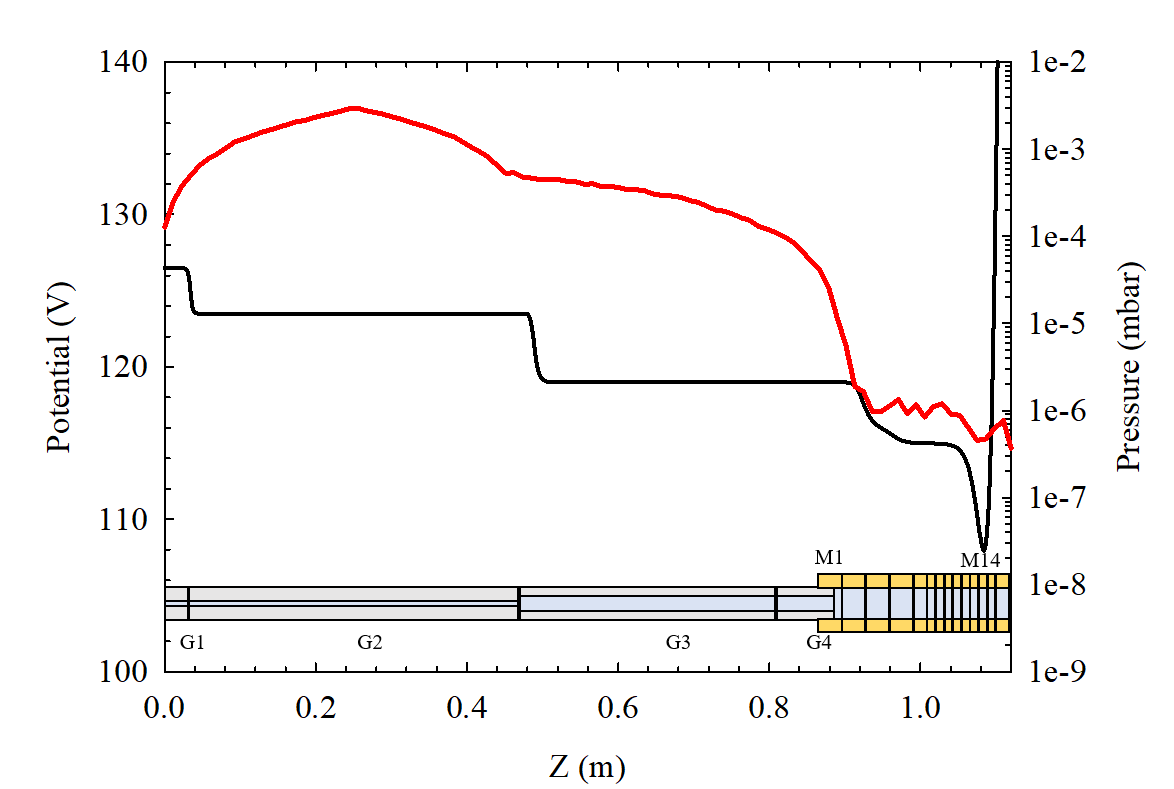}
  \caption{Electrical potential generated by the electrodes (LHS y-axis) shown as a black line. Simulation of pressure distribution of nitrogen buffer gas (RHS y-axis) shown as a red line made using Molflow+ (\cite{kersevan_introduction_2009}). Also shown, a schematic diagram of the positron trap electrodes to indicate position.}
\label{fig:GCP}
\end{figure}

The trap has a conventional gas cell structure consisting of a gate electrode (G1), a long narrow electrode (G2) with a high pressure of nitrogen buffer gas (see figure \ref{fig:GCP}), and finally electrodes with a larger inner diameter and lower pressure (G3 and G4). Pumping holes were drilled into the ends of G2 and G3 to allow gas to escape, lowering the pressure more quickly in the subsequent sections. The vacuum tube is split into two parts by a divider after G4, this means that most of the trap gases are pumped away on the upstream side of the apparatus. 
Beyond G4 on the other side of the pumping divider, there is a multi-ringed electrode (MRE) trap consisting of 14 (M1-M14) electrodes of the same diameter (42~mm) but varying lengths (see figure \ref{fig:GCP}, the narrowest electrodes are 10~mm wide). These create the final trapping potential well where positrons are stored before extraction to the Cusp trap. A typical trapping potential and the locations of the electrodes which create it, is also shown in figure \ref{fig:GCP}. The gas cell is constructed from aluminium, the MRE from oxygen free copper coated with a gold layer. Each electrode has an RC filter mounted inside the vacuum system to reduce electronic noise (by $\sim 100$ dB in the range $1-30$~MHz) picked up from the local environment. Positrons were extracted from the trap using a switching box that sends a fast pulse (fall time < 20~ns) to the confining electrode (M14) to open the well. \par 

%\begin{figure}
%\centering
%\begin{circuitikz}[scale=0.70] \draw
%%(0,0) to[R=200 <\kilo\ohm>, o-] (5,0)
%(5,0) to[R=200 <\kilo\ohm>] (10,0)
%(6,0) to[C=200 <\pico\farad>,*-] (6,-2) node[ground]{};

%\draw(9,0) to[C=200<\pico\farad>,*-] (9,-2) node[ground]{};

%\draw
%(1,4) to[empty diode] (9,4) 
%(9,2) to[empty diode, *-*] (1,2)
%(1,0) -- (1,4) 
%(9,0) -- (9,4) 
%;
%\end{circuitikz}
%\caption{\label{fig:circuit} Diagram of the filters, voltage from the power amplifier is supplied to the left hand input and the filtered output on the right supplied to the electrode.}
%\end{figure}

\begin{figure}
  \centering
  \includegraphics[scale = 0.5]{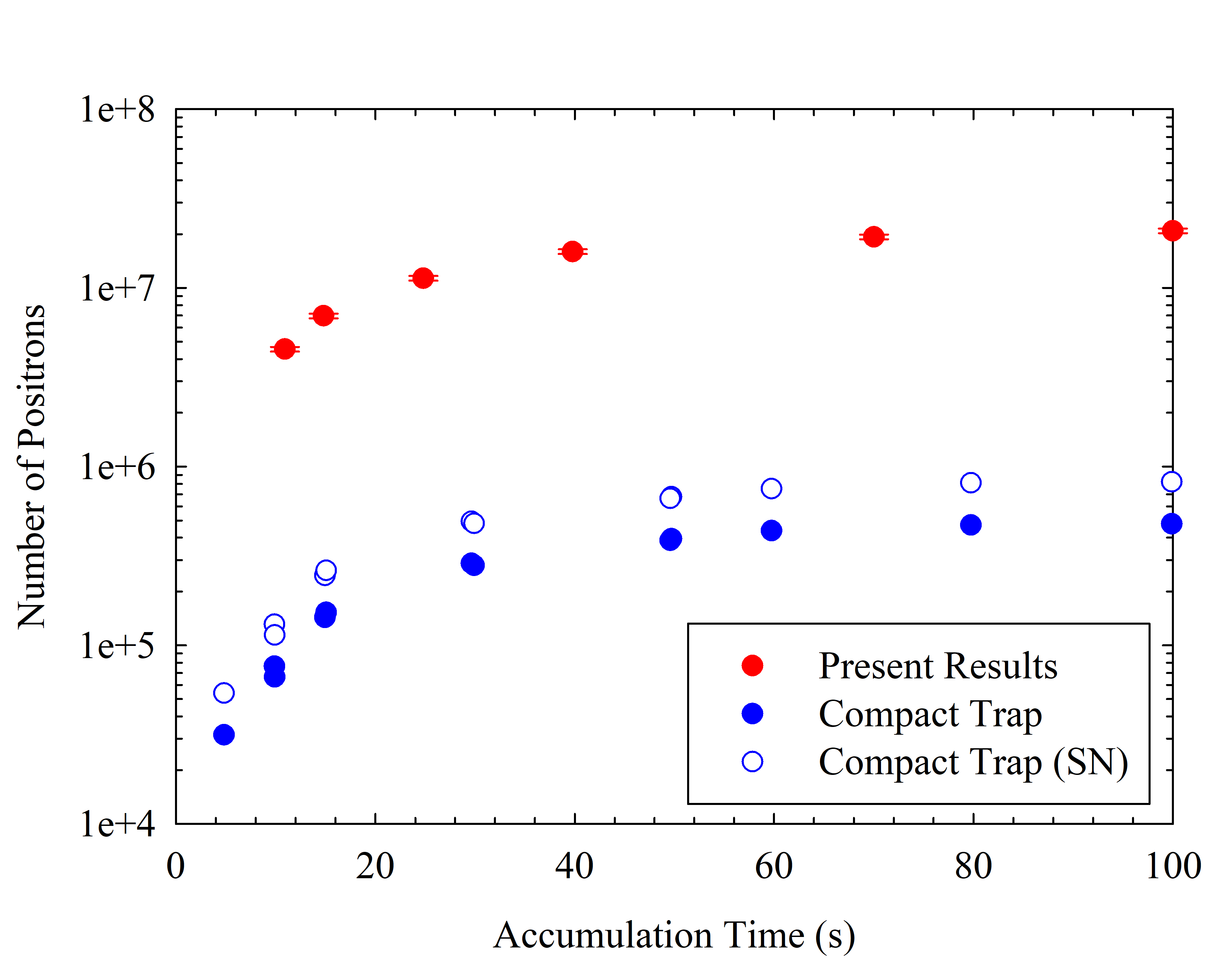}
  \caption{Number of positrons vs accumulation time. The present results are shown in red, previous results using the compact trap (\cite{fujii_positron_nodate}) are shown in blue. The hollow blue points show the compact trap data normalised by the difference in source activity for direct comparison. Error bars shown on red points are smaller than the symbols, these show the standard error.}
\label{fig:trapresults}
\end{figure}

A plot of positron number vs accumulation time is shown in figure \ref{fig:trapresults}. Results from the trap described above are shown as solid red points, results using the compact trap described by \cite{imao_positron_2010} are shown as solid blue points. For comparison, results from the compact trap, corrected by the different source activity are shown as blue hollow points. The present results show a trapped lifetime of 40~s and a peak trapping efficiency defined by the ratio of positrons entering the trap to those leaving, of $(17.4 \pm 1.8)\%$. The present trapped number is approximately 25 times higher than achieved with the compact trap. \par

\section{Positron accumulation}
\begin{figure}
  \centering
  \includegraphics[scale = 0.3]{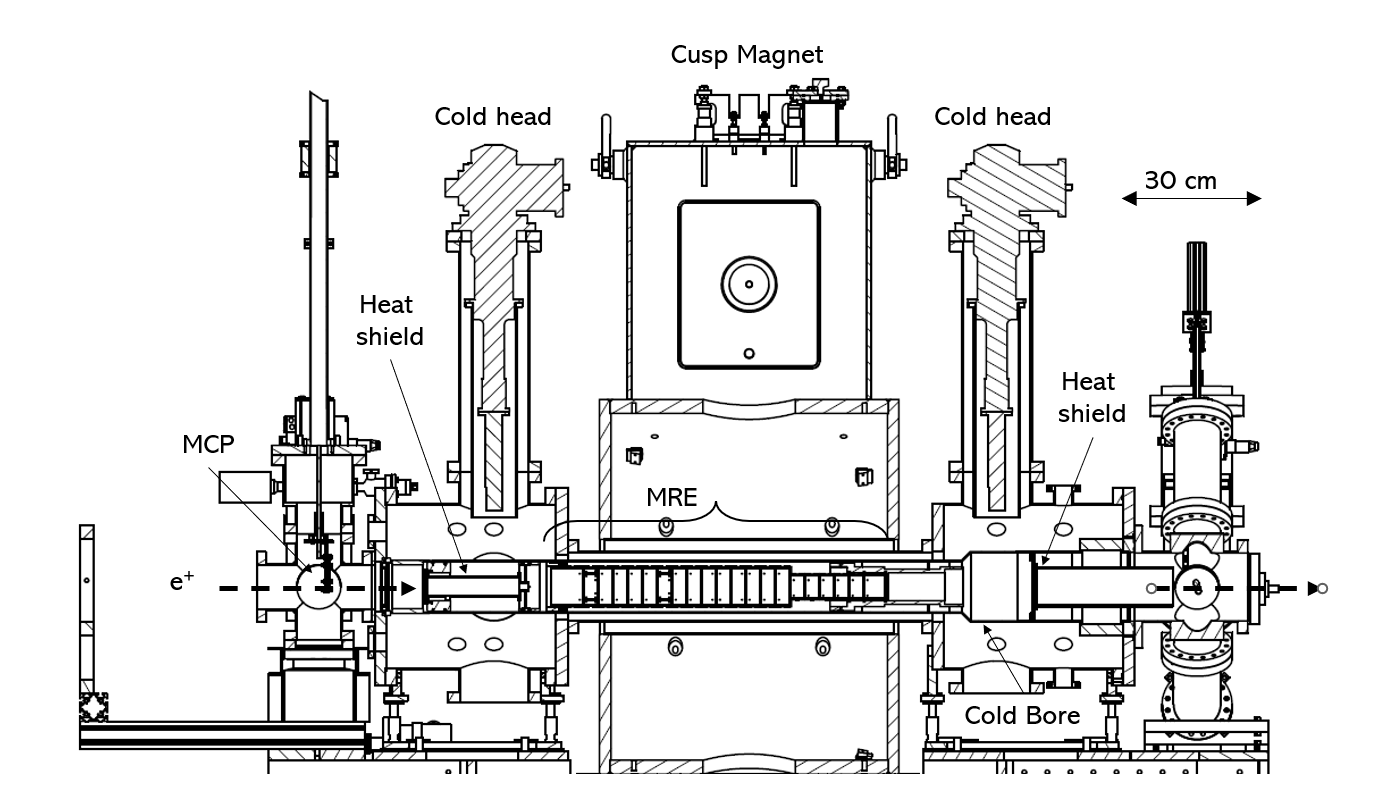}
  \caption{Scale drawing of the Cusp trap.}
\label{fig:cusp}
\end{figure}

During typical operation, the positron trap is filled for 40~s collecting approximately 15~million particles. Multiple bunches from the positron trap are accumulated in the Cusp trap to increase the number available for mixing. A scale drawing of the Cusp trap as used by 
 \cite{kolbinger_measurement_2021} is shown in figure \ref{fig:cusp}. Positrons are transferred via a magnetic conduit which is operated in a pulsed mode and has a magnetic field strength of up to 1000~gauss. Despite the 3~m length and two 90$^\circ$ turns, it is possible to achieve transport efficiencies of ($90\pm5$)~\% which is determined by comparing the number leaving the trap to those arriving at the Cusp trap MCP. As discussed above, the positron trap uses nitrogen buffer gas whereas the Cusp trap must have excellent vacuum for long antiproton lifetimes. To lower the contamination rate, gate valves along the transfer line are opened 5s before the pulse is extracted from the trap and closed shortly after.\par
 Positrons enter from the left hand side of the arrangement and are stored in the upstream section of the MRE. The Cusp MRE shown here consists of 21 electrodes made of oxygen free copper coated with gold, held in thermal contact with the cold bore using CuBe springs and mechanical wedges which can be tightened after the installation of the trap; the electrodes reach a temperature of approximately 15~K. Heat shields coated with a black nickel layer are located at either end of the trap to stop blackbody radiation from room temperature regions entering the MRE. An MCP and phosphor screen arrangement are located at the upstream entrance of the trap on a linear manipulator which can be used to image and count the number of particles extracted from the trap. 

\begin{figure}
  \centering
  \includegraphics[scale = 0.5]{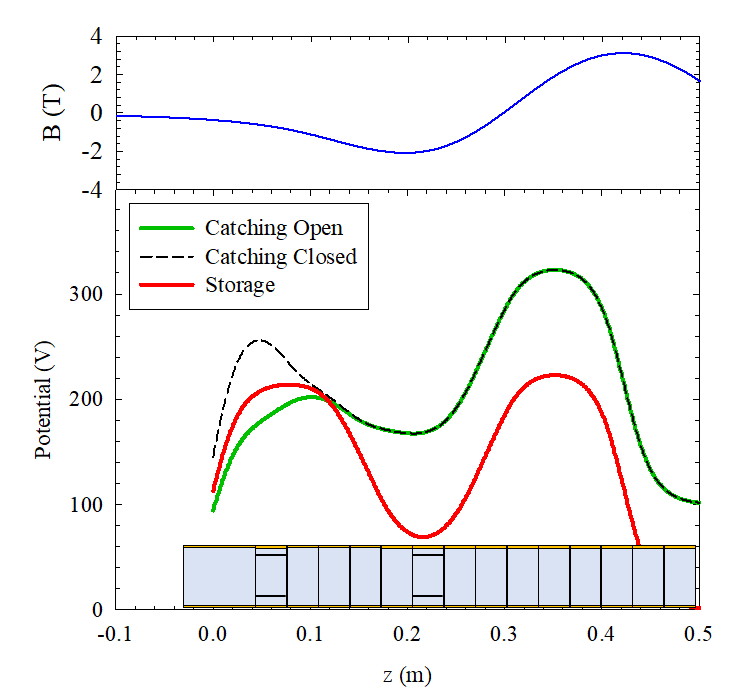}
  \caption{Magnetic field produced by the coils of the cusp magnet and the electrical potentials used when catching positrons from the trap. The location of the MRE electrodes are indicated the 2nd and 7th electrodes are segmented.}
\label{fig:cuspBElec}
\end{figure}

\begin{figure}
  \centering
  \includegraphics[scale = 0.5]{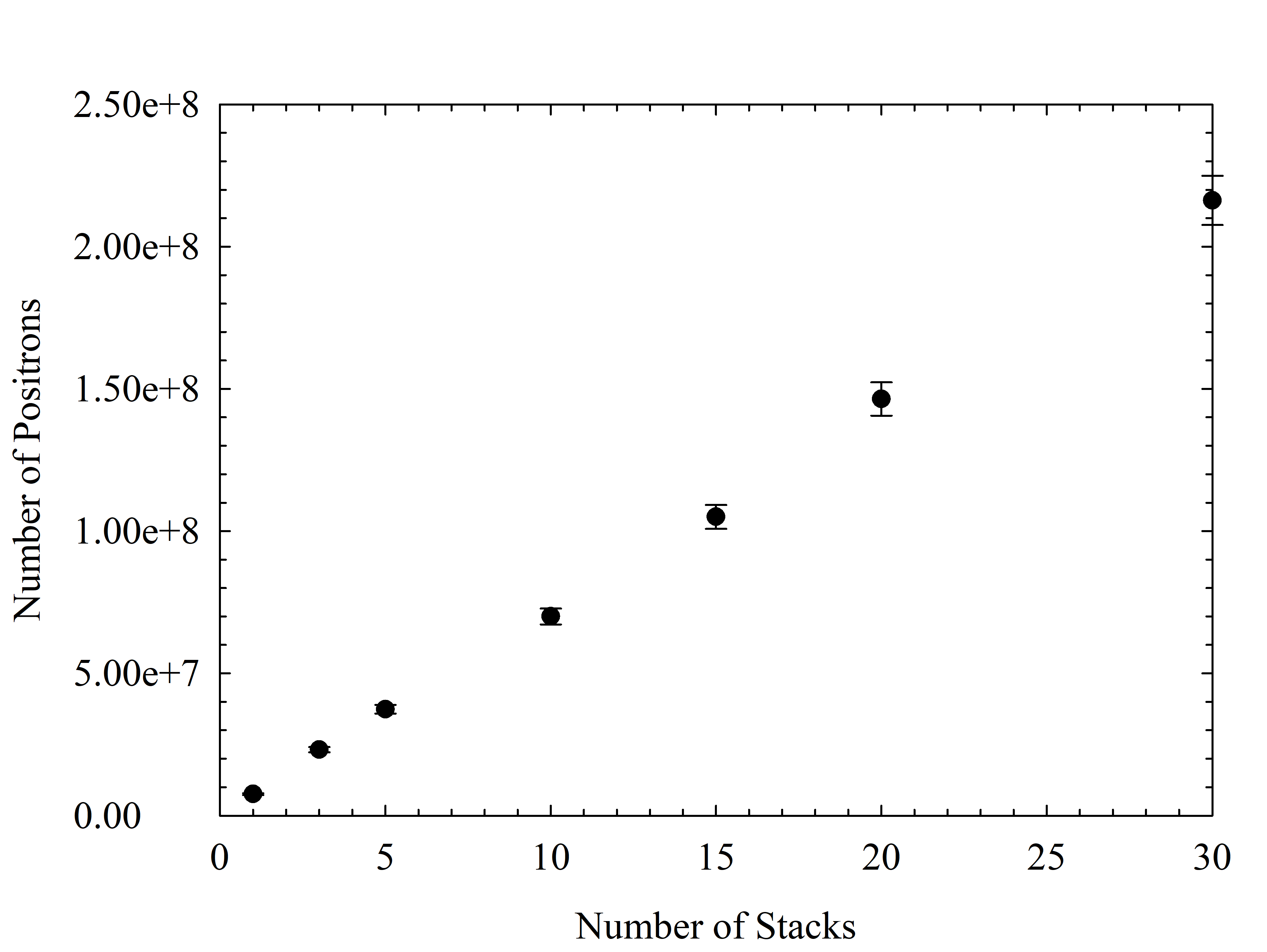}
  \caption{The number of positrons stored in the Cusp trap as a function of filling time.}
\label{fig:cuspfill}
\end{figure}

Figure \ref{fig:cuspBElec} shows the magnetic field produced by the Cusp magnet (blue line upper plot), the potentials used during the catch and store operation, and the position of the MRE electrodes. For each subsequent stack of positrons, the catching open potential (green solid line) is applied. Once the stack is inside the trap, the catching potential is closed (black dashed line). While the next stack is prepared, the particles are moved to the storage potential (red line) and a rotating wall electric field is applied to the segmented electrode to compress the plasma. The timing for this catching operation is controlled by a SRS-DG645 unit (Stanford Research Systems) which allows nanosecond level control over the triggering of the fast pulser to open the Cusp (fall time $\sim 20~ns$). The open catching potential is applied for $t_{open}\sim900$~ns during transfer. 
The number of positrons accumulated vs filling time is shown in figure \ref{fig:cuspfill}. Each stack takes approximately 50~s to fill the trap and perform transfer. The catching efficiency is typically ($50\pm5$)~\% depending on the vacuum conditions within the Cusp trap. The reproducibility in the number of positrons is approximately 5\%.\par
The optimum parameters for the rotating wall were found experimentally to be a sweep in the frequency from $f=7.0$~to~$7.5$~MHz during the filling time and peak-to-peak voltage of 5~V. Using the MCP and phosphor screen arrangement on the upstream side of the Cusp, a diameter of 1.6~mm is measured when the plasma is extracted. During mixing cycles, approximately 30~million positrons were used, which corresponds to a density of $\sim 2\times 10 ^{14}$~m$^{-3}$, assuming a uniform magnetic field across the trapping region.
\section{Discussion} 
There are still limitations associated with this system, firstly due to the use of a 6000~gauss field around the positron trap, a relatively high field (500~gauss) is required around the source region to prevent magnetic mirroring. Although mirroring can be overcome by increasing the beam energy, the magnetic field ratio increases the longitudinal energy spread making trapping behind the barrier at G1 more difficult. This may explain why the peak trapping efficiency is slightly lower than one would expect with two nitrogen buffer gas collisions.\par 
The use of the existing superconducting magnet also has the problem of the built in cold bore. It is impossible to operate the magnet without the cold bore in operation as LHe is consumed at an extremely high rate due to the proximity to the magnet. The LHe circulation loops used to cool the cold bore reach temperatures lower than 20~K meaning they cryo-pump nitrogen gas, however any instability in the flow rate of LHe around the cooling loops results in significant out gassing with pressures in the system increasing from the typical operating level of $1\times10^{-6}$~mbar to $1\times10^{-3}$~mbar or more, resulting in lost operational time. \par  
There is no noticeable effect of using a rotating wall electric field in the positron trap to increase the lifetime or the accumulated number. This may have been due to the lack of cooling gas which is omitted to improve vacuum during the transfer of particles to the Cusp. The magnetic field of 6000~gauss is expected to provide some cyclotron cooling, albeit with a lower rate than that of a cooling gas (\cite{greaves_inward_2000}).\par %for example compare $\Gamma_{c} = 0.09$~s$^{-1}$ and $\Gamma_{\textrm{SF}_6} = 3$s$^{-1}$  (at a pressure of $2\times10^{-8}$~torr from ).\par 
As the Cusp trap acts as the accumulator for the system, positrons which are produced when it is not available are wasted (e.g. when performing a mixing cycle or catching antiprotons). The need to transfer directly from the positron trap to the Cusp results in significant contamination of the Cusp trap vacuum (in operation, a 30~K warm up cycle of the Cusp cold bore is performed before each antiproton shift to ensure a long enough lifetime). It is not possible to pump out the gas before each transfer and replace it for the next cycle while maintaining an efficient duty cycle. \par 

\section{Conclusions} 
The results presented above show a filling rate corresponding to 4\% of the DC beam or 170,000~positrons~s$^{-1}$ which represents a rate improvement by a factor of ($52\pm3)$ over the previous compact design (\cite{fujii_positron_nodate}). The ability to quickly fill the Cusp trap with positrons has allowed more mixing experiments to take place during limited antiproton beam times which resulted in the first observation of an antihydrogen beam (\cite{kuroda_source_2014}) 2.7~m from the production region and the first measurements of the beam properties (\cite{kolbinger_measurement_2021}).\par 
Although this system is a great improvement and has helped to produce valuable results, the limitations discussed above must be considered in relation to the goal of producing an antihydrogen beam. The next steps to enhance performance and reliability will be to replace the superconducting magnet and its attached cold bore, used by the positron trap with a water cooled normal conducting magnet and room temperature bore. To improve upon the contamination of the Cusp trap and reduce positron waste, a 3$^{\textrm{rd}}$ stage accumulator will be constructed to allow transferring the positrons needed for mixing to be transferred in a single pulse from a region with good ($1\times10^{-8}$~mbar or lower) vacuum. \par

\section{Acknowledgements}
This work was supported by the Austrian Science Fund (FWF) Grant Nos. P 32468, W1252-N27, and P 34438; the JSPS KAKENHI Fostering Joint International Research Grant No. B 19KK0075; the Grant-in-Aid for Scientific Research Grant No. B 20H01930; Special Research Projects for Basic Science of RIKEN; Università di Brescia and Istituto Nazionale di Fisica Nucleare; and the European Union's Horizon 2020 research and innovation program under the Marie Sklodowska-Curie Grant Agreement No. 721559.

\bibliographystyle{jpp}
\bibliography{ASACUSA, JPPPositron}
\end{document}